\documentclass[conference]{IEEEtran}
\usepackage[utf8]{inputenc}
\usepackage{graphicx}
\usepackage{float}
\floatstyle{plain}
\newfloat{Code}{!ht}{myc}
\usepackage{microtype}
\usepackage{rotating}
\usepackage[breaklinks=true]{hyperref}
\usepackage{graphicx,xcolor}

\usepackage{listings}
\usepackage{multirow}
\lstdefinestyle{rdf}{numberblanklines=true, morekeywords={}}
\lstdefinestyle{sparql}{numberblanklines=true, morekeywords={SELECT,WHERE FILTER, GROUP BY, IN, AS}}

\usepackage{footnote}
\usepackage{placeins}
\usepackage{booktabs} 	
\usepackage{tabularx}
\usepackage{array}
\usepackage{subfigure}
\newcolumntype{R}{>{\raggedleft\arraybackslash}X}
\newcolumntype{L}{>{\raggedright\arraybackslash}X}
\newcolumntype{C}{>{\centering\arraybackslash}X}

\usepackage{eso-pic}
\usepackage{lipsum}
\AddToShipoutPicture{%
     \AtTextLowerLeft{%
         \put(-5,-30){\fbox{Presented at the 14th IEEE International Conference on Semantic Computing – Resource Track}}%
     }%
}

\date{}
\usepackage{tikz}
\usetikzlibrary{arrows,shapes,positioning,shadows,trees}
\newcommand*\circled[1]{\tikz[baseline=(char.base)]{
            \node[shape=circle,draw,inner sep=2pt] (char) {#1};}}

\begin{document}

\title{Towards an RDF Knowledge Graph of Scholars from Early Modern History}

\author{\IEEEauthorblockN{Jennifer Blanke}
\IEEEauthorblockA{Herzog August Library\\
Wolfenbüttel, Germany\\
blanke@hab.de}
\and
\IEEEauthorblockN{Thomas Riechert}
\IEEEauthorblockA{Leipzig University of Applied Sciences (HTWK)\\
Leipzig, Germany\\
thomas.riechert@htwk-leipzig.de}}

%\author{
%Jennifer Blanke\inst{1} \and
%Thomas Riechert\inst{2}
%\institute{Herzog August Bibliothek, Germany
%\and
%Leipzig University of Applied Sciences (HTWK), Germany
%\url{http://pcp-on-web.htwk-leipzig.de}
%}
%}

%\authorrunning{Blanke et al.}

\maketitle

\begin{abstract}
The use of Semantic Web Technologies supports research in the field of digital humanities. In this paper we focus on the creation of semantic independent online databases such as those of historical prosopography. These databases contain biographical information of historical persons. We focus on this information with an interest in German professorial career patterns from the 16th to the 18th century. In that respect, we describe the process of building an Early Modern Scholarly Career RDF Knowledge Graph from two existing prosopography online databases: the Catalogus Professorum Lipsiensium and the Catalogus Professorum Helmstadiensium. Further, we provide an insight in how to query the information using KBox to answer research questions.
\end{abstract}

\section{Introduction}
\label{sec:intro}

Semantic Web technologies brought semantics to several domains. 
One of these domains is historical research on online prosopography databases.
In this context, the process for building a research ontology consists of using historical expertise and knowledge engineering methods in parallel. 
It covers the database layer, the application
layer as well as the research interface layer of the Heloise Common Research Model
(HCRM)~\cite{thomasriechert2016collaborative}. 
Researchers start by exploring available external databases.
To that extent, queries are formally defined by SPARQL and can be used to access online
databases available through endpoints, distributed data management  frameworks such as KBox~\cite{marx2017kbox} or by URI-dataset indices such as WIMUQ~\cite{valdestilhas2018my}. 
By definition, SPARQL queries can be used to extract relevant concepts and properties for research vocabularies as well as be used to materialize relevant data into the envisaged research ontology.
This workflow enables researchers to re-build the research ontology at any time, as long as the syntax and semantics of the datasources are not getting changed.
The usage of a common vocabulary alleviates the problem of future data inconsistencies.
Additionally, the effort of exploring new
databases can be minimised as SPARQL can be used as a common vocabulary.
To that extent, there are many working groups targetting a common vocabularies such as the Data for History.\footnote{\url{http://www.dataforhistory.org}}

Although the Semantic Web technologies have contributed with vocabularies and tools to publish a huge variety of data on the web, 
most of its potential could only be achieved by integrating it.
That is the case of information of German historical scholars that are collected in different formats by each of the universities in Germany.

In this work, we discuss the different aspects of the Linked Data Life Cycle including the challenges and results of building a domain-specific 
RDF Knowledge Graph (KG) of Scholars and Scholarship from the 16th to the 18th century by gathering information from different online databases.
It describes the initial effort of fusing the Catalogus Professorum Lipsiensium\footnote{\url{https://research.uni-leipzig.de/catalogus-professorum-lipsiensium/}} and Catalogus Professorum Helmstadiensium\footnote{\url{http://uni-helmstedt.hab.de}}  into the Professor Career Patterns KG and enriching its content with data from Wikidata, DBpedia and Deutsche Nationalbibliothek by establishing \texttt{owl:sameAs} links among the professor's instances.
The datasets choice is given due to their heterogeneity and completeness nature.
The Catalogus Professorum Lipsiensium has information such as office and family relations.
The Professorum Helmstadiensium has extended this information with data about courses, students and their writing exams (Qualification Documents).
The Deutsche Nationalbibliothek has detailed information about the professors' publications.
The Wikidata and DBpedia are used to augment the KG with information such as religion, pictures, influencers, short bios, and alma maters.

The research aligns itself with the Digital Humanities for being interdisciplinary in
nature and for combining classic historiographical research methods with Semantic Web technologies in order to enable research on scholarly career.
%The research method uses classic prosopographic research questions which address a significant lacuna in the field of historical research.
The interdiciplinary research is conducted through the HCRM for cross-project research in the field of academic history.
We apply RDF standards for interlinking historical facts and to describe the vocabulary in a formal manner.
The interlinking process comprises the vocabulary and instance aligments as well as the quality control.
Finally, we present the evolution of the KG and clarify how the KG can be used to conduct histrorical research by using SPARQL queries.
The main contributions of this paper are: (1) an ontology and (2) a dataset of Early Modern scholars and scholarship, as well as (3) a detailed description of the methodological approach.
The \autoref{tab:resources} outlines the resources described in this work. 
It comprises (1) an ontology and (2) a dataset of Early Modern scholars available by \href{https://creativecommons.org/licenses/by-sa/4.0/}{CC BY-SA 4.0} license.

\begin{table*}
    \centering
    \begin{tabular}{r|l|l}
    \hline
    Resource & URL & License \\
    \hline\hline
         Ontology & \url{https://github.com/pcp-on-web/ontology} & \href{https://creativecommons.org/licenses/by-sa/4.0/}{CC BY-SA 4.0} \\ \hline
         Dataset  &  \url{https://gitlab.imn.htwk-leipzig.de/emarx/pcp-on-web/tree/master/dataset} & \href{https://creativecommons.org/licenses/by-sa/4.0/}{CC BY-SA 4.0} \\
         \hline
         \\
    \end{tabular}
    \caption{Resources described in this paper.}
    \label{tab:resources}
\end{table*}

\begin{figure*}
   \centering
   \includegraphics[width=0.85\textwidth]{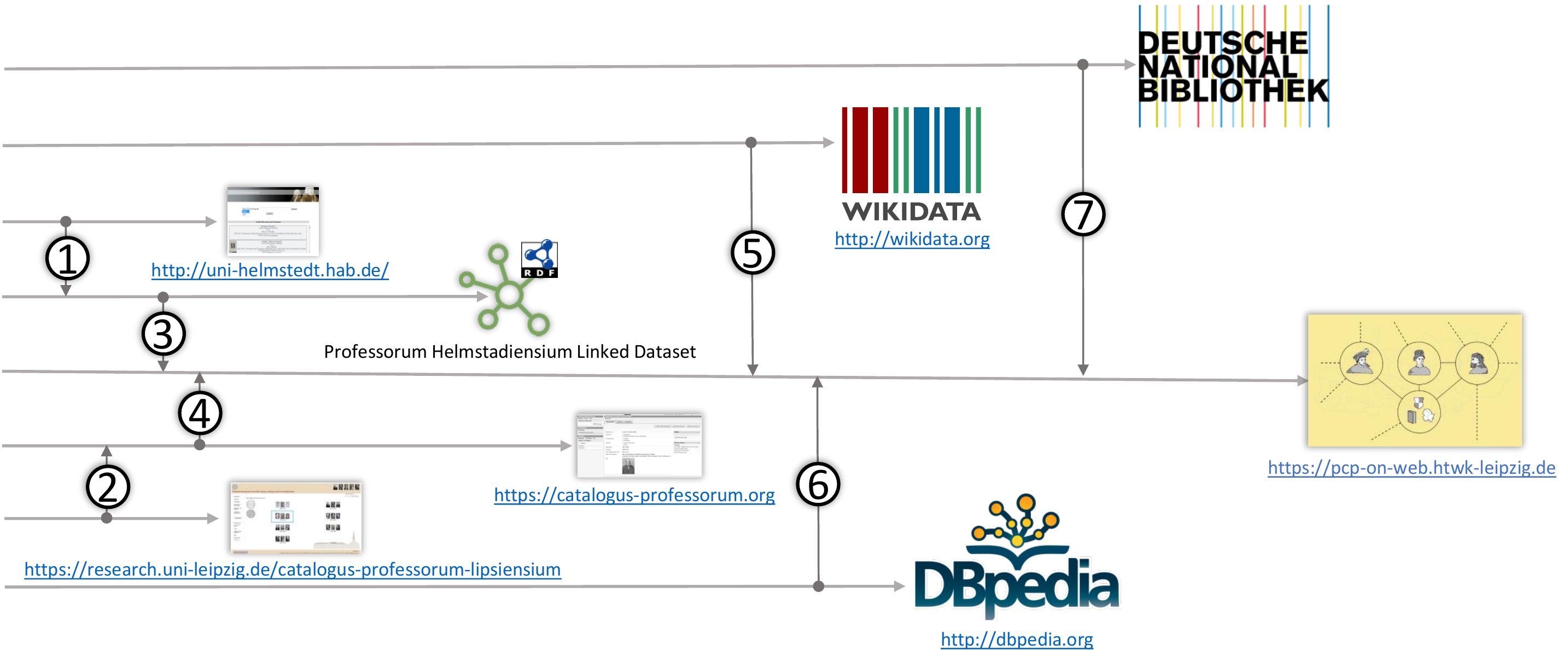} 
   \caption{Dataflow for creating the Professorial Career Patterns KG.}
   \label{fig:dataflow}
\end{figure*}

\section{Dataset}

The Professorial Career Patterns (PCP) RDF Knowledge Graph is composed of data extracted from different data sources.
\autoref{fig:dataflow} provides the dataflow of the data contained in the KG.
The data from the Catalogus Professorum Lipsiensium and the Catalogus Professorum Helmstadiensium was first extracted from the original databases and converted to RDF (\protect\circled{1} and \protect\circled{2}) in previous works.
The datasets were then processed and fused into the PCP KG (\protect\circled{3}-\protect\circled{4}) unther a common ontology,  described in \autoref{sec:ontology}.
The content of the KG is then enriched by interlinking the data from Wikidata (\protect\circled{5}), DBpedia (\protect\circled{6}) and the Deutsche Nationalbibliothek (\protect\circled{7}) through \texttt{owl:sameAs} links.
In the following sections we discuss the fusing, interlinking, evolving and exploring processes.

\section{Ontology}
\label{sec:ontology}

\begin{figure*}
   \centering
   \includegraphics[width=0.8\textwidth]{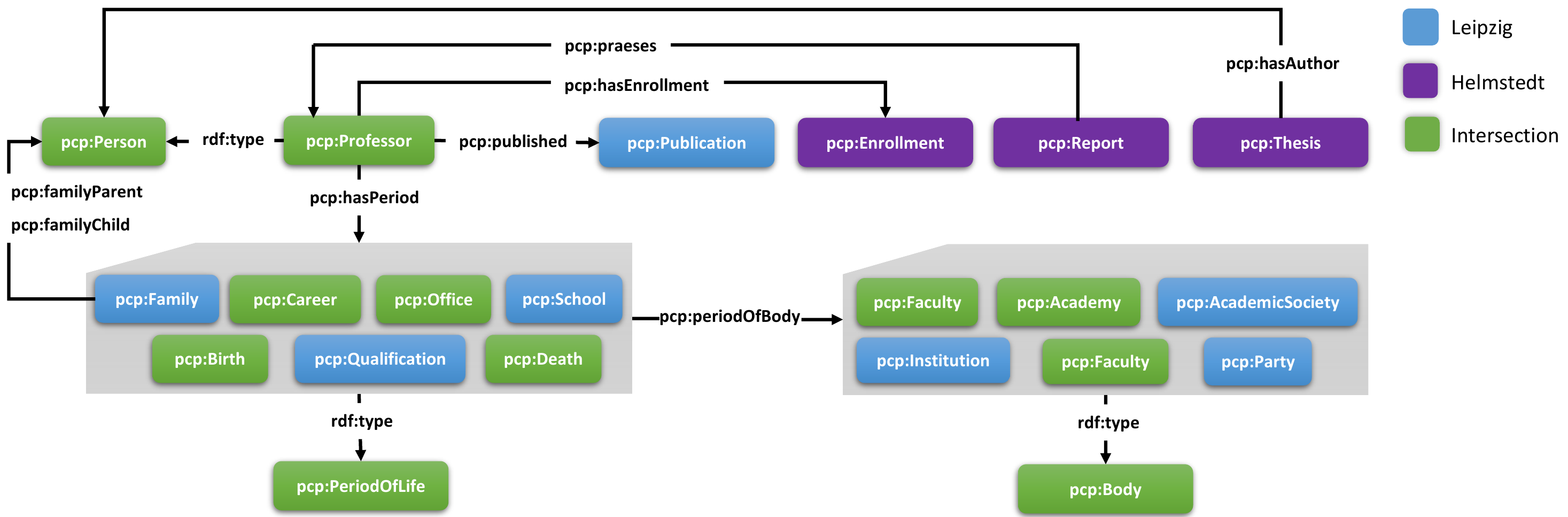} 
   \caption{An overview of the content fused in the Professorial Career Pattern KG. The blue classes are the content fused from Lipsiensium Catalogue. The purple classes are the content extracted from Helmstadiensium Catalogue. The green classes are the content available in both KGs.}
   \label{fig:content}
\end{figure*}

The ontology published in this work is bilingual, it contains label and concept descriptions either in English and German. 
The bilingual support enables transcultural transfer of specific Early Modern German scholarly concepts and simplifies data consumption at a later stage.
In  its  current  version,  the PCP Knowledge Graph describes \texttt{1185}  classes  and  \texttt{2099} properties.
An overview of the PCP-on-Web ontology is shown in \autoref{fig:content}.
The ontology is built around the class \texttt{pcp:Person} and its subclass \texttt{pcp:Professor}.
Following, we describe some of the ontology's classes.

\paragraph{Scholars} A prosopographical catalog gives basic information about persons, such as their name, birth and death data, as well as  data about their academic achievements.
The PCP-on-Web ontology allows to describe qualifications, publications, thesises, annual reports and more of the persons' academic and personal CVs.
One of the key properties in the KG is the PND using the Geimeinsame Normdatei (GND) of the German National Library. 
The GND can be used to identify historical persons.
Projects such as \texttt{PND/BAECON}\footnote{\url{https://old.datahub.io/dataset/pndbeacon}} enable to interlink other databases using the PND identifier.

\paragraph{Period of Life} To support a finegrained representation of different periods within the life of a person we introduced the concept \texttt{pcp:PeriodOfLife}, which is associated with a person through the properties \texttt{pcp:hasPeriod}. 
The ontology supports the modeling of different periods of life such as \texttt{pcp:Career}, \texttt{pcp:Office} (e.g. dean, rector), \texttt{pcp:Study}, \texttt{pcp:Qualification} (e.g. dissertation, and habilitation), \texttt{pcp:School}, \texttt{pcp:Birth} and \texttt{pcp:Death}. Each of these period of life subclasses contain different properties which are used to describe a particular instance in more detail. However, all inherit the delimiting properties \texttt{pcp:from} and \texttt{pcp:to} used to specify the period.
Different periods of life of the same person can overlap, e.g. the \texttt{pcp:Family} usually overlaps with other periods.

\paragraph{Body}. This class is used to describe relations among persons and organizations during a specific life time period (\texttt{pcp:PeriodOfLife}). Examples of bodies are the classes \texttt{pcp:Academy}, \texttt{pcp:AcademySociety}, \texttt{pcp:Faculty}, \texttt{pcp:Party}, \texttt{pcp:Institution}.
A person can belong to different bodies.

\paragraph{Family} Family relations are represented through the class \texttt{pcp:Family}. 
Instances of the class \texttt{pcp:Person} are then related to an instance of the \texttt{pcp:Family} class using the following properties: \texttt{pcp:familyChild}, \texttt{pcp:familyAdoptiveChild}, \texttt{pcp:familyFosterChild}, \texttt{pcp:familyParent}, \texttt{pcp:familyCohabitant}.

\paragraph{Academy} To add information related to the academic life of the scholars, the dataset contain calsses such as \texttt{pcp:Enrollment}, \texttt{pcp:Report}, \texttt{pcp:Thesis}, \texttt{pcp:Faculty} and even \texttt{pcp:Course}.
The dataset also includes porperties such as \texttt{pcp:lecturer}, \texttt{pcp:praeses} and \texttt{pcp:respondent}.
These metadata allows, for instance, to know the lectures given by a professor, who were his students, and how many thesis he has advised.

\section{Fusing}
This section describes the process of fusing the Catalogus Professorum Lipsiensium and the Catalogus Professorum Helmstadiensium into the Professorial Career Patterns KG.
It uses the following namespaces:
\paragraph{\textbf{\texttt{helmstedt}}} for Catalogus Professorum Helmstadiensium;
\paragraph{\textbf{\texttt{leipzig}}} for Catalogus Professorum Lipsiensium, and; 
\paragraph{\textbf{\texttt{pcp}}} for Professorial Career Patterns research ontology. \autoref{tab:properties} and \autoref{tab:classes} gives an overview of the properties and classes fused in this process.

\subsection{Vocabulary Aligment}

The first step of the interlinking process is vocabulary alignment.
This task was performed by a team of a computer scientist and a historical researcher, as a specialist on the domain specific databases.
In the process, properties and classes with same names or without corresponding counterpart were automatically shifted to the new namespace \url{http://purl.org/pcp-on-web/ontology#} (see \autoref{fig:content}), most properties/classes fit the criteria.
The remaining properties required a manual evaluation and further modification, discussed in Section \ref{sec:quality}. (Notice, other than Lipsiensium the Catalogus Professorum Helmstadiensium contains additional information about student enrollments and published thesises.)
The full task description can be found at \url{https://github.com/pcp-on-web/dataset/wiki/Data-sets-fusion:-task-description}.

\begin{table}
    \centering
    \begin{tabular}{r|r|r}
    \hline
    Operator & Leipzig & Helmstedt \\
    \hline\hline
         Joint  &  \multicolumn{2}{c}{21} \\
         \hline
         Disjoint &  51 & 35 \\\hline
         Union     & 72 & 56 \\\hline\hline
    \end{tabular}
    \caption{Overall Joint and Disjoint properties fused into the Professorial Career Patterns KG.}
    \label{tab:properties}
\end{table}

\begin{table}
    \centering
    \begin{tabular}{r|r|r}
    \hline
    Operator & Leipzig & Helmstedt \\
    \hline\hline
         Joint  &  \multicolumn{2}{c}{16} \\
         \hline
         Disjoint &  23 & 5 \\\hline
         Union     & 39 & 21 \\\hline\hline
    \end{tabular}
    \caption{Overall Joint and Disjoint classes fused into the Professorial Career Patterns KG.}
    \vspace{-20pt}
    \label{tab:classes}
\end{table}

\subsection{Instance Aligment}

In this task, we use the Link Discovery Framework Limes~\cite{ngomo2011limes} to align instances from both databases, the Catalogus Professorum Lipsiensium and the Catalogus Professorum Helmstadiensium. The aim was to merge resource instances from both catalogues that refer to one and the same person, in order to find parallels and to enrich the prosopgraphical data. To achieve that goal, we applied the Limes framework using two distinct configurations. The first, using acceptance at \texttt{0.8}, and, the second at \texttt{0.5}. 
Both setups used instances for persons with the following attributes: name (\texttt{rdfs:label}), surname and forename. 
The algorithm used was the unsupervised version of ``wombat simple."
We also tried a full match using the acceptance rate of \texttt{1}. But there was no instance of a person in both data sets that exactly match one another.

\begin{itemize}
    \item \textbf{First Setup} The configuration with the highest acceptance rate (\texttt{0.8}).
    That means that the Person instances \texttt{leipzig:heinrichmatthiasheinrichs} and \texttt{helmstedt:13084} achieved a similarty score of \texttt{0.81}.
    Following, we manually checked the two instances, concluding that they both refer to different persons (see Listing \ref{match08}).
    As can be seen in Listing \ref{heinrichmatthiasheinrichs} and \ref{helmstedt}, the instances cannot refer to the same person as either surnames and forenames are different.

\begin{lstlisting}[caption={Similarity measure between leipzig:heinrichmatthiasheinr\\ichs and helmstedt:13084.},label=match08]
leipzig:heinrichmatthiasheinrichs helmstedt:13084	0.8164965809277261
\end{lstlisting}

\begin{lstlisting}[caption={Properties surname, forname and label from the instance \\ leipzig:heinrichmatthiasheinrichs in Leipzig Professor's catalog \\database.},label=heinrichmatthiasheinrichs]
leipzig:heinrichmatthiasheinrichs leipzig:surname "Heinrichs" ; leipzig:forename "Heinrich Matthias" ; rdfs:label "Heinrich Matthias Heinrichs" .
\end{lstlisting}

\begin{lstlisting}[caption={Properties surname, forname and label from the Person's \\instance helmstedt:13084 in Helmstedt database.},label=helmstedt]
helmstedt:13084 rdfs:label "Andreas Heinrich Matthias" . helmstedt:13084 helmstedt:forename "Andreas Heinrich" . helmstedt:13084 helmstedt:surname "Matthias" .
\end{lstlisting}

\item \textbf{Second Setup} The configuration with the lowest acceptance rate of \texttt{0.5} encountered \texttt{1683} instances with possible alignment.
The manual analysis of the result carried out by the historian did not find any matching instance.

\end{itemize}

\subsection{Quality Control}
\label{sec:quality}
After interlinking we perform a quality check in the ontology performed by two historian data experts.
The quality check was designed to fix nomenclature errors and to enhance the properties and classes descriptions.
Surprisly, the Helmstedt ontology concepts did not contain labels and descriptions, requiring their creation.
Thus, an historian manually performed the ontology's metadata creation in two languages German and English for 35 properties (see \autoref{tab:properties}) and 5 classes (see \autoref{tab:classes}).
Few properties and classes were renamed to standardize the use of English and German.
Property's descriptions were enhanced to better describe their usage.
Some of the common errors were:

\begin{itemize}
    \item multi-lingual labels (e.g. \texttt{pcp:hasMatrikel});
    \item different naming patterns (e.g. \texttt{pcp:surname\_lat} instead of \texttt{pcp:latinSurname}), and;
    \item wrong labeling concept (e.g. \texttt{pcp:lecture} became \texttt{pcp:lecturer}).\footnote{The lecture (to give a lecture) was misconceived by the lecturer (the person who gives the lecture).}
\end{itemize}

The full task description can be found at \url{https://github.com/pcp-on-web/dataset/wiki/Instance-Matching:-Link-Discovery}.

\section{Interlinking}

Among many information that can be found on the PCP KG, there is the GND information of the professors available through the \texttt{pcp:gnd} property.
The main idea of the interlinking process is to use the GND to link relevant data available in other KGs and use this data for enrichment.
This section describes PCP KG professors interlinking process with their respective instances on DBpedia, Wikidata and the Deutsche Nationalbibliothek (DNB) by the use of \texttt{owl:sameAs} links as well as the data extraction.

\subsection{The GND standardization}
Although the GND is available in the PCP KG professor's instances, one it is not standardized.
The Catalogus Professorum Lipsiensium uses the GND number while the Catalogus Professorum Helmstadiensium uses the Deutsche Nationalbibliothek (DNB) GND URL. 
To overcome this issue, we replace the URL by the GND number. 
This can be done by extracting the GND number from the URL. 
The Deutsche Nationalbibliothek GND URL is a composition of the GND's namespace and the GND number e.g. in the URL \url{https://d-nb.info/gnd/118755951}, the GND namespace is \texttt{https://d-nb.info/gnd} and the GND number is \texttt{118755951}.

After the standardization the process of interlinking using \texttt{owl:sameAs} links as well as the extraction of Wikidata, DBpedia and DNB professors' instances was set.
The DNB data can be interlinked by reconstructing the DNB URL using the professors' GND number followed by the suffix ``/about/lds.'' 
Although the RDF data describes the GND URL (e.g. \url{https://d-nb.info/gnd/118755951}), it can only be accessed with the “about/lds” suffix addition (e.g. \url{https://d-nb.info/gnd/118755951/about/lds}).

Once the DNB URL is reconstructed, it can also be used to interlink and extract data from Wikidata and DBpedia.
That's because either Wikidata and DBpedia professor's instances contain reference to their respective GND URL. 
In Wikidata, the GND URL is referenced by the property \url{http://www.wikidata.org/prop/direct/P227} and in DBpedia by the property \texttt{owl:sameAs}.
By using these properties, it is possible to lookup for DBpedia and Wikidata professors containing the GND URL in their corresponding respective properties.

%that are \texttt{owl:sameAs} the PCP professors' DNB URL as well as for Wikidata professors that have the PCP professors' DNB URL referenced by the property \url{http://www.wikidata.org/prop/direct/P227}.

\subsection{Extractor}

We also conducted the extraction from the relevant subsets of DNB, Wikidata and DBpedia datasets.
To this aim, a lazy-extraction approach was designed. 
The approach receives a list of GNDs and a SPARQL query template. 
It performs the instance extraction one by one, therefore lazy-extraction. 
The aim is to avoid server timeouts and errors by executing simple and fast SPARQL queries. 
The approach is open-source and is publicly available at \url{https://github.com/pcp-on-web/scholar.extractor}.

The Wikidata, DNB, and DBpedia extracted data is publicly available at the dataset Github page. 
There, users can report issues or subscribe to receive update notifications.
It is also possible to query it locally using KBox~\cite{marx2017kbox} (see \autoref{lst:subsets}) to simplify the sharing and querying.
The \autoref{tab:subsets} give the total number of classes and properties for each of the KG subsets.

\begin{lstlisting}[caption={Querying different Professorial Career Patterns subgraphs.},label=lst:subsets]
Java -jar kbox-v0.0.1.jar -kb "http://purl.org/pcp-on-web/dbpedia,http://purl.org/pcp-on-web/wikidata,http://purl.org/pcp-on-web/dnb" -sparql "Select * where {?s ?p ?o}" -install
\end{lstlisting}

\begin{table}
    \centering
    \begin{tabular}{r|r|r}
    \hline
    Dataset & \#Properties & \#Classes \\
    \hline\hline
         Helmstedt & 56 & 21 \\ \hline
         Leipzig  &  72 & 39 \\ \hline
         DNB  &  81 & 8 \\ \hline
         DBpedia  &  246 & 1136 \\ \hline
         Wikidata  & 1682 & 6 \\ \hline\hline
         Total & 2099 & 1185 \\ \hline
    \end{tabular}
    \caption{Professorial Career Patterns subgraph statistics.}
    \label{tab:subsets}
\end{table}

\section{Evolving}

\begin{figure}
   \centering
   \includegraphics[width=\columnwidth]{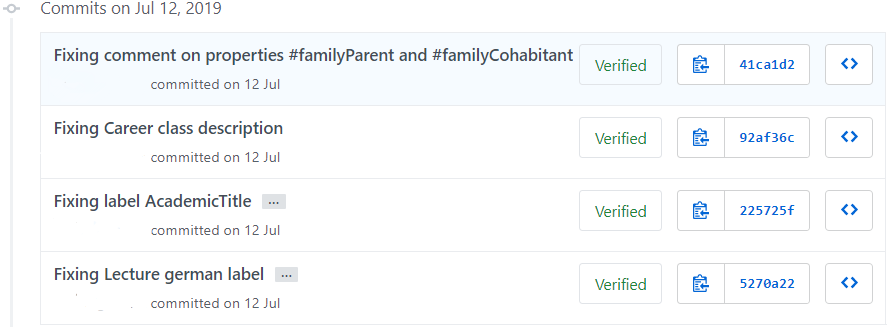} 
   \caption{Example of the alterations on the Professorial Career Patterns ontology, publicy available at \url{https://github.com/pcp-on-web/ontology/commits/master}.}
   \label{fig:evolving}
   \vspace{-20px}
\end{figure}

Due to the distributed character of Web of Data, approaches that provide versioning and provenance play a central role.
It is important to track the provenance of data at any step of a process involving possible changes of a database (e.g., creation, curation, linking).
It provides a good basis for mechanisms to track down and debug the origin of errors and improve processes.
Envisioning an approach to support a collaborative database curation and research made \texttt{QUIT}~\cite{arndt2019decentralized} a natural choice.
\texttt{QUIT} enables access to provenance-related metadata pertaining to the KG and provide all functionalities of a version control system using Git.
\autoref{fig:evolving} depicts a list of alterations performed by commits in the ontology repository.
It is possible to visualize and explore the changeset as well as follow the ontology and KG evolution.

\section{Exploring}

\begin{figure}
   \centering
   \includegraphics[width=0.6\columnwidth]{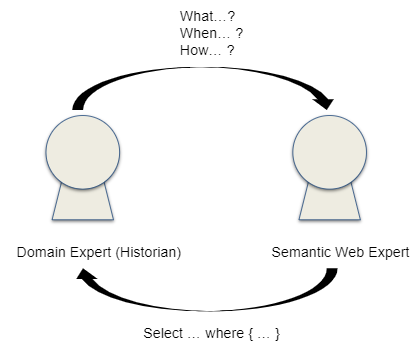} 
   \caption{Interaction between domain and Semantic Web expert.}
   \label{fig:interaction}
   
\end{figure}

Due to the flexibility of the SPARQL language and the lack of a practical approach to bridge the knowledge between the data and the Semantic Web experts, we apply an interactive approach to enable researchers to explore the data.
The data expert provides a question in natural language to the Semantic Web expert who formulates the SPARQL query.
The SPARQL query result is then checked by the data expert providing research insights and error analyses (Figure \ref{fig:interaction}).
The Listing \ref{question} and \ref{kbox} provide an example of a question and its respective SPARQL query.
To make the interlinked database accessible to other researchers we use KBox.
The database and ontology are both available under the Knowledge Name (KN)\footnote{Knowledge Name is a reference to the KG in KBox.} http://purl.org/pcp-on-web/dataset and http://purl.org/pcp-on-web/ontology (see Listing \ref{kbox}).

\begin{lstlisting}[caption={Properties surname, forname and label from the Person instance helmstedt:13084 in Helmstedt database.},label=question]
Give me the amount of qualification documents with professors as praeses, arranged by year and faculty.
\end{lstlisting}

\begin{lstlisting}[caption={Listening properties from the merged ontology using KBox. In this example, the namespace declaration was omitted for the purpose of simplification.},label=kbox, basicstyle=\scriptsize]
java -jar kbox-v0.0.1.jar -sparql "select (count(?doc) as ?docN) ?faculty ?year 
where { 
    ?doc pcp:praeses ?professor. 
    ?doc a pcp:QualificationDocument. 
    ?professor a pcp:Professor . 
    ?doc pcp:faculty ?faculty. 
    ?doc pcp:date ?docDate. 
    bind (year(?docDate) as ?year ). 
} group by ?faculty ?year order by asc(?year) asc(?faculty)" -kb "http://purl.org/pcp-on-web/dataset,http://purl.org/pcp-on-web/ontology,"http://purl.org/pcp-on-web/dbpedia,http://purl.org/pcp-on-web/wikidata,http://purl.org/pcp-on-web/dnb" -install
\end{lstlisting}

\section{Conclusion}
\label{sec:conclusion}

In this paper, we described the interlinking process of two prosopographical databases in order to conduct research on the research question of Early Modern scholarly career patterns.
We gave insight into the different steps which are involved in the vocabulary and databases curation as well as quality control.
We further discussed the data evolution and an exploratory research method engaging data domain and Semantic Web experts.

\bibliographystyle{abbrv}
\bibliography{refs}

\begin{thebibliography}{1}

\bibitem{arndt2019decentralized}
N.~Arndt, P.~Naumann, N.~Radtke, M.~Martin, and E.~Marx.
\newblock Decentralized {C}ollaborative {K}nowledge {M}anagement using {G}it.
\newblock {\em Journal of Web Semantics}, 54:29--47, 2019.

\bibitem{marx2017kbox}
E.~Marx, C.~Baron, T.~Soru, and S.~Auer.
\newblock Kbox—{T}ransparently {S}hifting {Q}uery {E}xecution on {K}nowledge
  {G}raphs to the {E}dge.
\newblock In {\em 2017 IEEE 11th International Conference on Semantic Computing
  (ICSC)}, pages 125--132. IEEE, 2017.

\bibitem{thomasriechert2016collaborative}
T.~Riechert and F.~Beretta.
\newblock Collaborative research on academic history using linked open data: A
  proposal for the heloise common research model.
\newblock {\em CIAN-Revista de Historia de las Universidades}, 19(0), 2016.

\bibitem{ngomo2011limes}
M.~A. Sherif, A.-C.~N. Ngomo, and J.~Lehmann.
\newblock Wombat--a generalization approach for automatic link discovery.
\newblock In {\em European Semantic Web Conference}, pages 103--119. Springer,
  2017.

\bibitem{valdestilhas2018my}
A.~Valdestilhas, T.~Soru, and M.~Saleem.
\newblock More complete resultset retrieval from large heterogeneous rdf
  sources.
\newblock In {\em Proceedings of the 10th International Conference on Knowledge
  Capture}, pages 223--230, 2019.

\end{thebibliography}

\end{document}